\documentclass{elsart}

\usepackage{amssymb}
\usepackage{graphicx}

\begin{document}

\begin{frontmatter}

\title{Coevolution of agents and networks: Opinion spreading and community disconnection}

\author{Santiago Gil \and Dami\'an H. Zanette\thanksref{Z}}
\address{Centro At\'omico Bariloche and Instituto Balseiro,
8400 Bariloche, R\'{\i}o Negro, Argentina}

\thanks[Z]{Also at Consejo Nacional de Investigaciones
Cient\'{\i}ficas y T\'ecnicas, Argentina.}

\begin{abstract}
We study a stochastic model for the coevolution of a process of
opinion formation in a population of agents and the network which
underlies their interaction. Interaction links can break when agents
fail to reach an opinion agreement. The structure of the network and
the distribution of opinions over the population evolve towards a
state where the population is divided into disconnected communities
whose agents share the same opinion. The statistical properties of
this final state vary considerably as the model parameters are
changed. Community sizes and their internal connectivity are the
quantities used to characterize such variations.
\end{abstract}


\end{frontmatter}

In a complex system, the spontaneous emergence of collective
non-equilibrium behaviour, such as coherent spatiotemporal
structures or synchronized dynamics, is driven by mechanisms which
involve both the interactions between the system's components and
external influxes \cite{ref1,ref2}. In segregation phenomena, a form
of self-organization well known to physicists and chemists, an
ensemble of interacting elements becomes split into subensembles
whose components share certain individual states. Segregation occurs
also in biological and social systems \cite{ref3,ref4,ref5}, where
it plays a crucial role in sustaining diversity at many levels
--cellular, functional, organizational, ecological, cultural. Though
it is usually associated with phase separation in space, segregation
not always takes place in the spatial domain. In human societies,
for instance, two or more subpopulations or communities may coexist
in the same geographical region and, yet, exhibit mutually excluding
cultural traits \cite{ref6}. With respect to those traits --which
may involve religious beliefs, professional or generational jargons,
artistic inclinations, etc.-- such communities are scarcely
interacting, and can be effectively considered as segregated  from
each other. A key mechanism in this kind of social segregation is
the feedback between the construction of agreement within a
community and the enhancement of distinctions with other
communities: specific traits become better established as
differences between communities develop and grow.

In the last years, physicists have been increasingly interested in
the dynamical and statistical modelling of complex systems of
biological and sociological inspiration as populations of agents
whose interaction patterns are described through graphs, or
networks. Much attention has been paid to dynamical processes
defined over quenched networks  \cite{ref2,ref7}, and to network
growth --both purely stochastic  \cite{ref8} or driven by adaptive
mechanisms  \cite{ref9}-- with emphasis in the statistical
properties of the resulting patterns. On the other hand, the
possible coevolution of the network structure and the dynamics
taking place over them seems to have been less studied (see,
however, Ref. \cite{ref9a}). This kind of coevolution is at the
basis of the feedback phenomenon pointed out above, where the
formation of internally homogeneous subpopulations and the weakening
of their mutual interactions enhance each other.

The aim of this letter is to present a very simple model for
coevolution of a population of agents and their interaction network.
The agents' dynamics is based on an elementary model of opinion
spreading \cite{ref10}. The population starts in a situation where
every agent is able to interact with any other agent, and evolves
towards a segregated state with disconnected communities.
Interactions between agents with similar or different opinions are
respectively favoured or penalized. In spite of the simplicity of
the evolution rules, the population can reach a variety of social
patterns, which range from splitting into two communities of similar
sizes and opposite opinions, to a single large community with
homogeneous opinion. Typical connectivities and opinion
distributions, resulting from the combined evolution of the network
and the agent states, change considerably with the control
parameters.

The system consists of $N$ agents situated at the nodes of a
network. Initially, the network is fully connected, so that any pair
of agents can potentially interact. The individual state of each
agent is assigned at random one of two possible values, say $+1$ or
$-1$, with equal probability. Individual states represent the
agents' opinions, which may eventually change by interaction between
agents connected by the network.

The evolution runs as follows. At each step, a pair of connected
agents is chosen at random from the whole population. If both
agents have the same opinion, nothing happens. Otherwise, with
probability $p_1$ either agent adopts the other agent's opinion,
so that the two opinions become identical. With the complementary
probability $1-p_1$, opinions are not changed. In this case,
however, the link between both agents is broken with probability
$p_2$, and the interaction network looses one of its edges. These
rules are successively applied until no further changes are
possible. Since network edges are irreversibly deleted by the
evolution, a frozen final state is eventually reached where,
generally, the network is split into disconnected subsets. Within
each of these communities all agents share the same opinion.

The frequencies of the individual events that drive the dynamics
depend on the probabilities $p_1$ and $p_2$. With respect to the
evolution of the system, these probabilities are independent
control parameters. The statistical properties of the final state,
however, are completely determined by the relative frequencies of
the different processes that effectively change the state of the
system. In other words, they depend on $p_1$ and $p_2$ through
certain algebraic combinations only. To realize this, let
$p_-(t)$ be the fraction of network links connecting agents with
different opinions. The probability that any agent changes its
opinion at a given step is $\pi_1(t)=p_-(t)p_1$, and the
probability that a connection is broken is
$\pi_2(t)=p_-(t)(1-p_1)p_2$. The sum $\pi(t)=\pi_1(t)+\pi_2(t)$
is the probability per evolution step that any change takes
place, and thus fixes an overall evolution time scale. If $p_1$
and $p_2$ vary in such a way that the ratios $q=\pi_1(t)/\pi(t)$
and $1-q=\pi_2(t)/\pi(t)$ are kept constant, such overall time
scale will change, but the relative frequency of the two
processes will be the same, giving rise to statistically
equivalent final states. Thus, the only independent combination of
the probabilities $p_1$ and $p_2$ relevant to the determination of
the final state is
\begin{equation}
q=\frac{\pi_1(t)}{\pi(t)}=\frac{p_1}{p_1+(1-p_1)p_2}.
\end{equation}

The behaviour at the two extreme values of $q$ is immediately
assessed. For $q=0$ (i.e. $p_1=0$), where no  opinion changes take
place, the final structure consists of two mutually disconnected
communities with similar sizes and opposite opinions. Internally,
each of them stays fully connected, so that the total number of
remaining connected pairs $R$ is close to $2(N/2)(N/2-1)/2 \approx
N^2/4$. For $q=1$ (i.e. $p_1=1$), on the other hand, interacting
agents with initially opposite opinions always reach consensus, so
that no interaction links are broken. At the final state, all
agents share the same opinion and the network is still fully
connected, with its $N(N-1)/2\approx N^2/2$ links intact.

\begin{figure}
\centering \resizebox{\columnwidth}{!}{\includegraphics{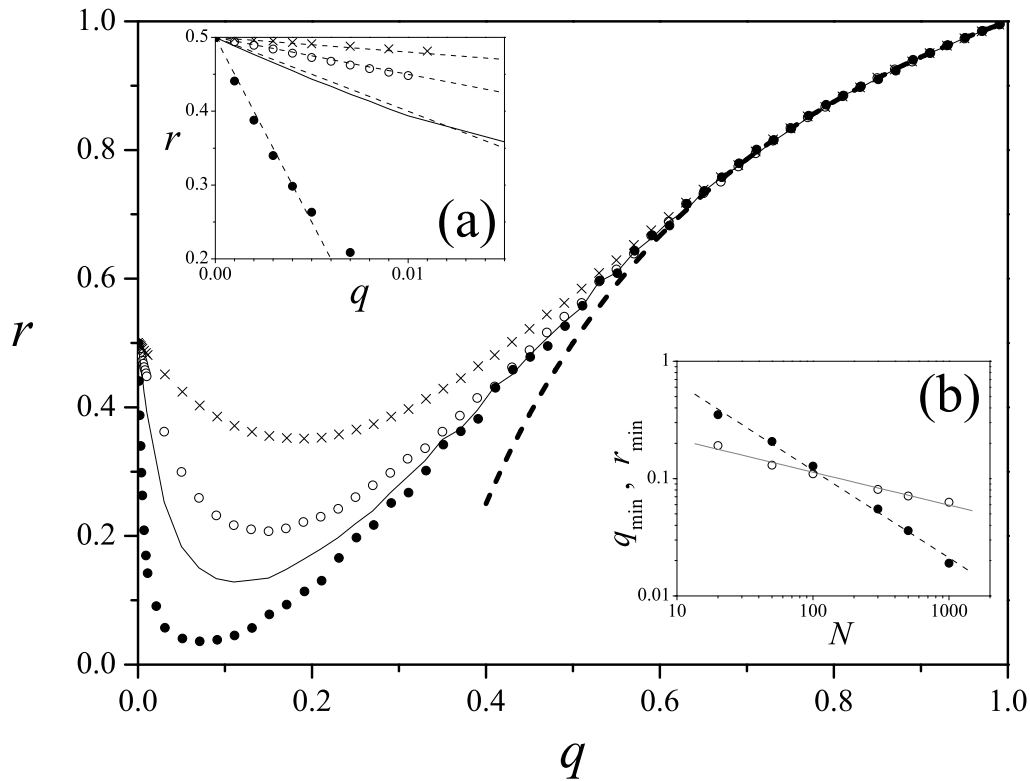}}
\caption{The  fraction $r$ of remaining links as a function of the
parameter $q$, for $N=20$ ($\times$), $50$ ($\circ$), $100$ (full
line, and $500$ ($\bullet$). The dashed line represent the
analytical approximation for large $q$ discussed in the text. Insert
(a): Close-up of the same data for small $q$. The dashed lines
represent the small-$q$ approximation discussed in the text. Insert
(b): The position $q_{\min}$ ($\circ$) and the depth $r_{\min}$
($\bullet$) of the minimum of $r$ as functions of $N$. Straight
lines are least-square fittings with slopes $-0.28$ (full line) and
$-0.75$ (dashed line).} \label{f1}
\end{figure}

These two limits suggest that the fraction of remaining connected
pairs, $r=2R/N(N-1)$, provides a first quantitative characterization
of the final state. Figure \ref{f1} shows $r$ as a function of $q$,
for systems of various sizes $N$. For each value of $q$, the
fraction $r$ has been obtained as the average over $500$ to $5000$
realizations of the initial condition and the evolution. Rather
unexpectedly, we find that $r$ reaches a minimum for an intermediate
value of $q$. The position $q_{\min}$ and depth $r_{\min}$ of this
minimum depend on $N$, and seem to tend to zero as the population
grows, as shown by the plot in insert (b). The network connectivity
at the minimum is considerably lower than at the extreme values
$q=0$ and $1$. For $N=100$, for instance, we have $r_{\min} \approx
0.13$, which implies an average connectivity of about $13$ links per
site. In this intermediate regime, thus, the network is poorly
connected and the population structure correspondingly degraded.

The minimum at $q_{\min}$ defines two regimes in the parameter $q$.
To the left ($q<q_{\min}$), as expected from the behaviour for
$q=0$, the population becomes split into two separate communities
with similar sizes and opposite opinions. As $q$ grows towards
$q_{\min}$, the internal connectivity of these communities
decreases. Simultaneously, several small separate clusters, each of
them containing just a few agents (typically, less than $5$ for
$N=500$), are also found at the final state.

\begin{figure}
\centering \resizebox{\columnwidth}{!}{\includegraphics{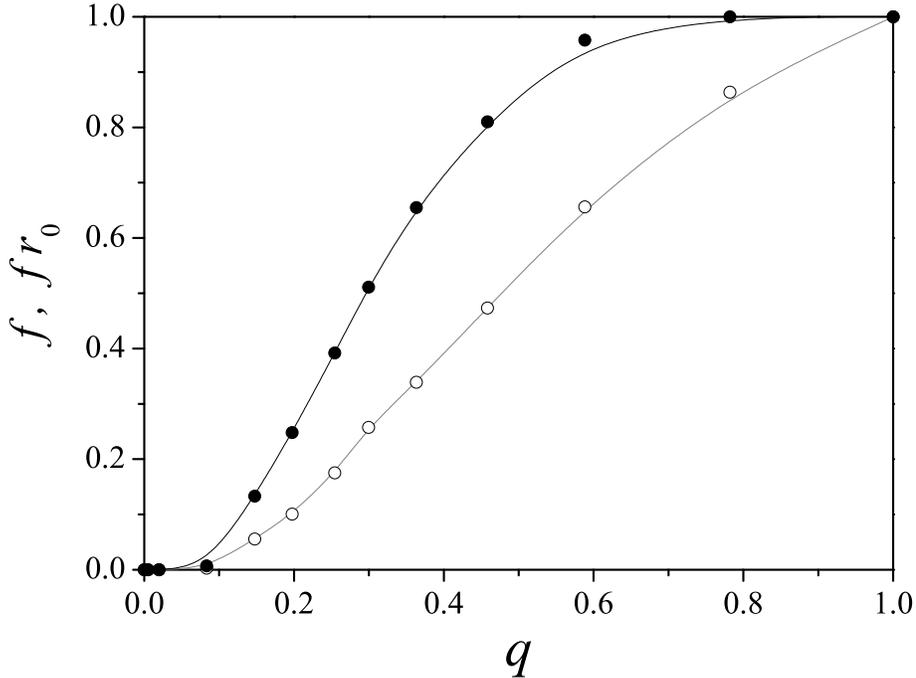}}
\caption{ The fraction $f$ of realizations ending in a single
community ($\bullet$), and the contribution $fr_0$ of those
realizations to the fraction of remaining links ($\circ$), where
$r_0$ is the average number of remaining links in a single-community
final state. Curves ar B\'ezier splines, added for clarity.}
\label{f2}
\end{figure}

To the right of the minimum, we still have realizations where the
population splits into disconnected communities. The two largest
communities always have opposite opinions, but their size is much
more variable than for $q<q_{\min}$. Moreover, the number and size
of small communities grow. At the same time, it becomes increasingly
frequent to find realizations where the whole population stays
connected into a single community. In these realizations, the final
network is not fully connected, but its connectivity is
significantly larger than in the cases where separate communities
are formed. As a matter of fact, it is the contribution of these
realizations which determines the growth of the fraction of
connected pairs $r$ for $q>q_{\min}$. As $q$ keeps growing, the
fraction of realizations with a single-community final state
increases and, eventually, all realizations end on such states, with
a homogeneous opinion all over the system. Figure \ref{f2} shows the
fraction $f$ of realizations where the population reaches a
single-community final state as a function of $q$ for $N=500$, and
the contribution of those realizations to the fraction of remaining
links.

\begin{figure}
\centering \resizebox{\columnwidth}{!}{\includegraphics{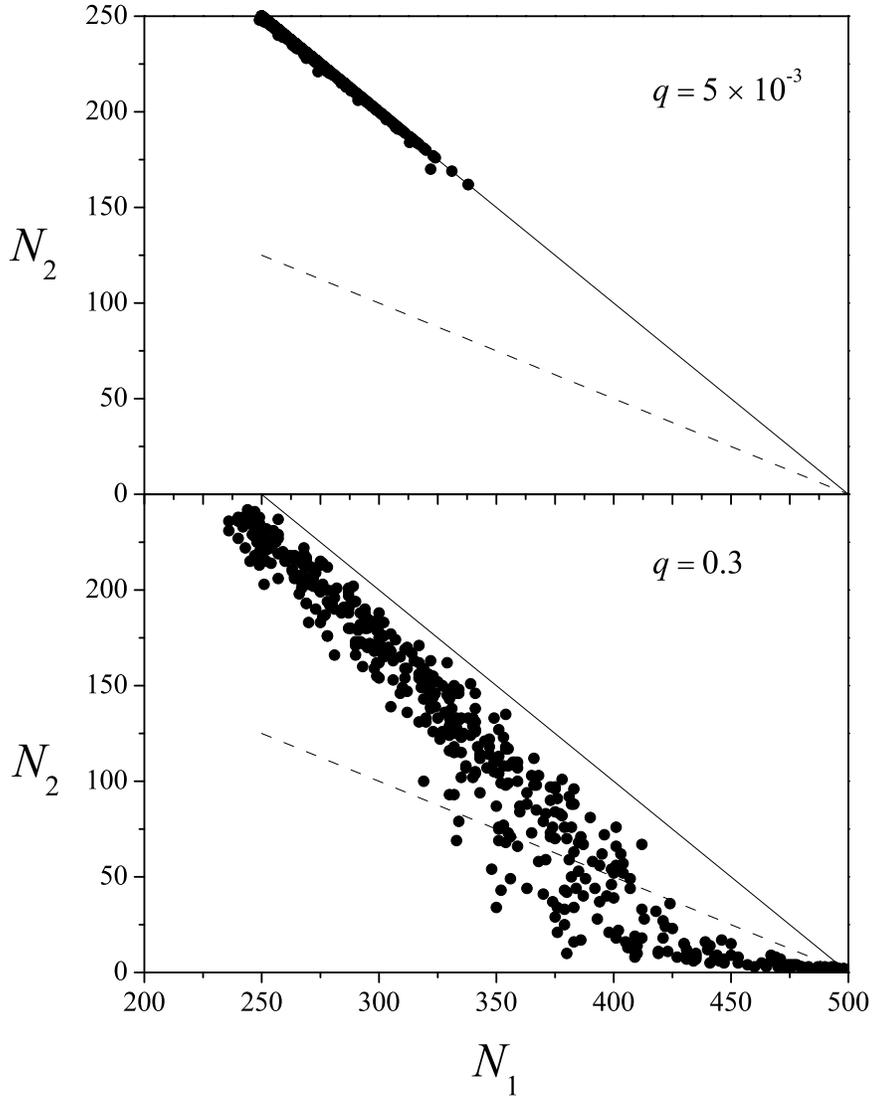}}
\caption{The size $N_2$ of the second largest community as a
function of the size $N_1$ of the largest community, for two values
of the control parameter $q$, in a series of $10^3$ realizations of
a system with $N=500$ agents. Each dot corresponds to a single
realization. The straight and the dashed lines represent,
respectively, the functions $N_2=N-N_1$ and $N_2=(N-N_1)/2$.}
\label{f3}
\end{figure}

In order to compare the regimes at both sides of the minimum, we
have chosen to study in more detail the final structure of a
population of size $N=500$ for two values of $q$ where the fraction
of remaining connected pairs $r$ attains similar levels, namely,
$q=5\times 10^{-3}$ and $0.3$ ($r\approx 0.27$; see Fig. \ref{f1}).
Figure \ref{f3} shows, for the two chosen values of $q$, the size of
the second largest community ($N_2$) as a function of the size of
the largest community ($N_1$), for $10^3$ realizations. Generally,
$N_2\neq 0$, except in realizations where the final state consists
of a single community, where $N_1=N$ and $N_2$ vanishes. The
straight full line stands for the function $N_2=N-N_1$, so that the
vertical distance from each dot to the line represents the size
$N_0=N-N_1-N_2$ of the population not included in the two largest
communities for the corresponding realization. The dashed line, in
turn, corresponds to $N_2=(N-N_1)/2$. Dots below this line represent
realizations where the second largest community is smaller than
$N_0$.

For $q<q_{\min}$, we find that the size of the largest community
reaches, at most, $N_1\approx 340 \approx 0.7 N$. Moreover, most
dots lie over the straight line, which implies that essentially
all agents are in one of the two largest communities ($N_0\approx
0$).  For $q>q_{\min}$, on the other hand, the largest community
can reach the maximum size $N_1=500=N$, indicating that some
realizations already correspond to single-community final states.
Now, all sizes between $N_1\lesssim N/2$ and $N$ seem to be
possible. Except in the close proximity of $N_1=N$, where the
largest community comprises practically the whole population, dots
are sensibly below the full line. The population $N_0$ not
belonging to the two largest communities is typically around $0.1
N$. Moreover, for most realizations where $N_1 \gtrsim 400 = 0.8
N$, this population is above the size of the second largest
community. In this situation, we can no longer properly speak of
splitting into two main communities. The final structure is in
fact closer to the single-community state, with the addition of
several small separated communities.

\begin{figure}
\centering \resizebox{\columnwidth}{!}{\includegraphics{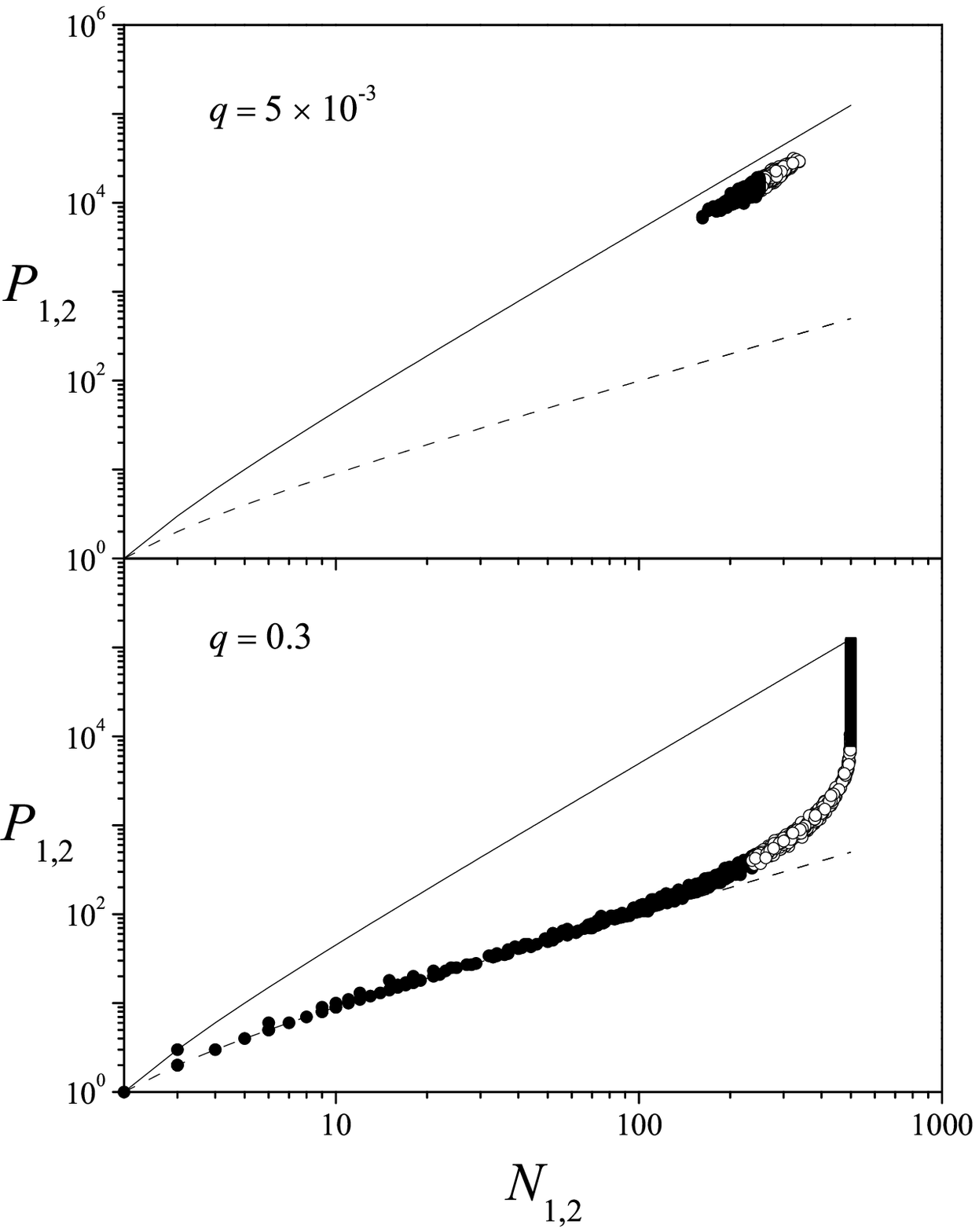}}
\caption{The number of remaining links $P_i$ within the largest
($\circ$) and the second largest ($\bullet$) community ($i=1,2$,
respectively), as a function of the community size $N_i$, for two
values of $q$, in a series of $10^3$ realizations of a system of
$N=500$ agents. Realization with single-community final states are
indicated by a different symbol ($\blacksquare$). The full curve in
each plot represents the function $P_{1,2}=N_{1,2}(N_{1,2}-1)/2$;
the dashed curve is $P_{1,2}=N_{1,2}-1$. } \label{f4}
\end{figure}

The distribution of internal connectivities in the resulting
communities is also strongly dependent on the parameter $q$. Figure
\ref{f4} shows the number of links $P_i$ within each community as a
function of the community size $N_i$, for the two largest
communities in each realization ($i=1,2$). Two different symbols
identify the largest ($\circ$) and the second largest ($\bullet$)
communities. A third symbol ($\blacksquare$) is used for
realizations with a single-community final state ($N_1=N$). The full
curve represents the function $P_{1,2}=N_{1,2}(N_{1,2}-1)/2$,
corresponding to fully connected networks. The dashed curve stands
for $P_{1,2}=N_{1,2}-1$, the minimum number of links in a connected
network.

For $q<q_{\min}$, the number of remaining links in each of the two
largest communities exhibits a limited dispersion between
realizations --just like their size $N_{1,2}$. As a result, all the
realizations are represented by dots in a rather compact cloud.
Within each community, roughly one half of the total possible
connections are actually present, $P_{1,2}\approx
N_{1,2}(N_{1,2}-1)/4$, so that the average connectivity per site is
about $N_{1,2}/2$. In this situation, thus, the internal
connectivity of both communities is quite high.

For $q>q_{\min}$, in contrast, the number of remaining links is
broadly distributed. For practically all sizes, the number of links
of second-largest communities is just above the minimum,
$P_{1,2}=N_{1,2}-1$, indicating a barely connected structure. The
average connectivity per site within these communities is, in fact,
close to one. The number of links in the largest communities, on the
other hand, are always considerably above the minimum. However, even
the better connected among these communities are still far from the
situation of full connectivity. For realizations where $N_1 \lesssim
N/2$, $P_1$ reaches at most $10 \% $ of its maximum value. The
number of links almost attains the maximum in some of the
realizations where the population stays connected in a single
community. These realizations, in fact, fill the gap between the
best connected communities in the case of split populations and
fully connected networks. Interestingly, while the above description
emphasizes the differences between the internal structure of
second-largest communities, largest communities, and
single-community populations, the fact that all the dots in the
lower panel of Fig. \ref{f4} seem to lie over a smooth curve suggest
that those different structures belong to a sole class. The internal
connectivity varies continuously as the community size changes.

Obviously, the coevolution of the distribution of opinions and the
interaction network creates strong correlations between the
individual states of those agent pairs that remain connected. These
correlations make an accurate analytical treatment of our model
particularly difficult. Some of the features found in the numerical
results, however, admit an analytical explanation under suitable
hypotheses. Let us begin by the regime of large $q$ where, as we
have seen, most realizations end in a single-community state. The
evolution towards a homogeneous opinion all over the population is
fully driven by fluctuations. In fact, the number $N_+$ of agents
with opinion $+1$, for instance, performs a random walk, $N_+\to
N_+\pm 1$, each step an opinion changes. Starting at $N_+\approx
N/2$, the process ends when $N_+$ reaches either $0$ or $N$. From
standard results on first-passage time distributions in random walks
\cite{ref11}, we find that the average number of active steps needed
to reach one of these extreme values is $S=N^2/4p_1$. An active step
is here defined as an evolution step where any change, either the
flip of an individual opinion or the deletion of a network link,
occurs. In an active step, a link is deleted with probability
$(1-p_1)p_2$. If we assume that this latter event is not correlated
with opinion changes, the total number of links deleted in $S$
active steps is $N^2(1-p_1)p_2/4p_1$. Taking into account that the
initial number of links is, approximately, $N^2/2$, we have
\begin{equation}
r\approx \frac{1}{N^2/2}\left[\frac{N^2}{2}- \frac{N^2}{4}
\frac{(1-p_1)p_2}{p_1}\right]= 1-\frac{1-q}{2q}.
\end{equation}
Note that this estimation for the fraction of remaining links is
independent of $N$. It is represented by the dashed curve in Fig.
\ref{f1}. We see that the agreement with numerical results is
excellent for large $q$. While neglecting correlations between link
deletion and opinion changes is not strictly correct --as a matter
of fact, links are deleted precisely in those steps where opinions
{\it do not} change-- for $q\approx 1$ deletion events are so rare
that such correlations can hardly accumulate into a relevant effect
before the final state is reached.

For small $q$, on the other hand, the final state always
corresponds to essentially two large communities with opposite
opinions. Since, in this limit, $p_1$ is small, most links
connecting agents with different opinions are deleted before any
significant number of opinion flips occurs. At any time, thus,
essentially all links connect agents with the same opinion, except
just after an opinion flip: when an agent's opinion changes, most
of its links are now connections with agents with the opposite
opinion. The total number $P_+$ of links connecting agents with
the same opinion remains constant until an opinion flip takes
place. In such event, the decrease in $P_+$ is given by the number
of links of the agent whose opinion is changing, which can
considerably vary among agents. As a kind of mean-field approach,
we assume that this number is proportional to the average
per-agent number of links connecting agents with the same opinion,
$2P_+/N$, so that $\dot P_+ =-2 \alpha p_1 P_+/N$. The derivative
is performed with respect to a time variable whose units are
active steps, and the heuristic factor $\alpha$ represents the
proportionality assumed above. The solution to this equation is
\begin{equation}
P_+(s) = \frac{N^2}{4} \exp \left( -\frac{2\alpha p_1}{N}
s\right),
\end{equation}
where $s$ is the number of elapsed active steps. Moreover, as
discussed above, the total number of links after $s$ active steps
is
\begin{equation}
P(s) = \frac{N^2}{2}-(1-p_1)p_2 s.
\end{equation}
The evolution ends when $P_+(s)=P(s)$, so that the fraction $r$ of
remaining links is determined by the equation
\begin{equation}
r=\frac{1}{2} \exp \left[ -\alpha N\frac{q}{1-q} (1-r)\right].
\end{equation}
Insert (a) of Fig. \ref{f1} shows fittings of the numerical
results for small $q$, obtained from this equation with
$\alpha=0.4$. The fact that $\alpha$ does not depend on the size
$N$ implies that $r$ depends on $q$ and $N$ through their product
$qN$. This explains, at least qualitatively, that the minimum in
$r$ shifts to smaller $q$ as $N$ grows. This shift is a direct
consequence of the different dependence on $N$ of the two
mechanisms which drive the system for small $q$. While at each
active step only one link can be deleted, the number of links
connecting agents with equal or different opinions can change in
a quantity of order $N$.

In summary, we have shown that our simple model of coevolution for a
population of agents and the underlying interaction network gives
rise to an interesting variety of population structures. The
evolution rules represent the spreading of a bivaluate opinion on a
network whose links can break when agents do not succeed at reaching
an opinion agreement. The final population structure consists,
typically, of a set of separate communities, each of them containing
agents with the same opinion. The resulting structures can be
divided into three main classes: (i) two internally well connected
communities with similar sizes and opposite opinions; (ii) a single
community containing all the population; and (iii) a well connected
community with, typically, more than half the population,
accompanied by a set of poorly connected smaller communities. In
this latter class, the two largest communities always have opposite
opinions.

While a clear boundary between the three classes cannot be
unambiguously drawn, class (i) on the one hand, and (ii) and (iii)
on the other, are observed within different ranges of the control
parameter $q \in [0,1]$. The cross-over region between these two
ranges moves towards $q=0$ when the population size $N$ grows, so
that the small-$q$ regime would disappear for $N\to \infty$. This
effect can be avoided if, when two agents fail to reach an opinion
agreement, not only their mutual connection but a given fraction of
their links to other agents is also broken. In this way, deletion of
links and opinion change would have the same dependence on the
system size. Preliminary numerical results of this extension of the
model have already been obtained, and will be published in a
forthcoming paper. Another extension, in the direction of making the
model more realistic, would allow for the creation of links between
disconnected agents. This process would drive the population to a
dynamic asymptotic state, independent of the initial structure,
where communities could form, aggregate, exchange agents, and
disappear, as known to happen in real social systems.


\begin{thebibliography}{00}

\bibitem{ref1} A. S. Mikhailov, {\it Foundations of Synergetics I. Distributed and Active
Systems} (Springer, Berlin, 1994).

\bibitem{ref2} S. C. Manrubia, A. S. Mikhailov and D. H. Zanette, {\it Emergence of Dynamical
Order. Synchronization Phenomena in Complex Systems} (World
Scientific, Singapore, 2004).

\bibitem{ref3} J. D. Murray, {\it Mathematical Biology} (Springer, Berlin,
1993).

\bibitem{ref4} J. Portugali, {\it Self-Organization and the City} (Springer, Berlin,
2000).

\bibitem{ref5} R. Axelrod, {\it The Complexity of Cooperation} (Princeton University Press, Princeton,
1997).

\bibitem{ref6} R. Dunbar, Ch. Knight and C. Power, eds., {\it The Evolution of Culture.
An Interdisciplinary View} (Rutgers University Press, New Brunswick,
1999).

\bibitem{ref7} D. J. Watts, Small Worlds (Princeton University Press, Princeton,
1999).

A. Barrat  and M. Weigt, Eur. Phys. J. B {\bf 13}, 547 (2000).

M. Ch\'avez, D. U. Hwuang, A. Amann, H. G. E. Hentschel and S.
Boccaletti, Phys. Rev. Lett. {\bf 94}, 218701 (2005).

\bibitem{ref8} A. L. Bar\'abasi and R. Albert, Science {\bf 286},
509 (1999).

R. Pastor Satorras, J. Rub\'{\i} and A. D\'{\i}az Guilera, eds.,
{\it Statistical Mechanics of Complex Networks} (Springer, Berlin,
2003).

\bibitem{ref9} D. van den Berg and C. van Leeuwen, Europhys. Lett.
{\bf 65}, 459 (2004).

\bibitem{ref9a} M. G. Zimmermann, V. M. Egu\'{\i}luz and M. San
Miguel, Phys. Rev. {\bf 69}, 065102(R) (2004).

G. C. M. A. Ehrhardt, M. Marsili and F. Vega Redondo, Emergence and
resilience of social networks: A general theoretical framework,
physics/0504124.

P. Holme and M. E. J. Newman, Nonequilibrium phase transition in the
coevolution of networks and opinions, physics/0603023.



\bibitem{ref10} M. Levy, H. Levy and S. Solomon, {\it Microscopic Simulation
of Financial Markets} (Academic Press, New York, 2000).

\bibitem{ref11} C. W. Gardiner, {\it Handbook of Stochastic Methods}
(Springer, Berlin, 1997).

\end{thebibliography}
\end{document}